\documentclass[prb,twocolumn,showpacs,preprintnumbers,amsmath,amssymb,superscriptaddress]{revtex4}

\usepackage{graphicx}
\usepackage{dcolumn}
\usepackage{bm}

\begin{document}


\title{Temperature - pressure phase diagram of the superconducting iron pnictide LiFeP}

\author{K. Mydeen} \email{kamal@cpfs.mpg.de}
 \affiliation{%
Max Planck Institute for Chemical Physics of Solids, 01187 Dresden, Germany
}%
\author{E. Lengyel}
\affiliation{%
Max Planck Institute for Chemical Physics of Solids, 01187 Dresden, Germany
}%
\author{Z. Deng} \affiliation{%
Institute of Physics, Chinese Academy of Sciences - Beijing, China
}%
\author{X. C. Wang} \affiliation{%
Institute of Physics, Chinese Academy of Sciences - Beijing, China
}%
\author{C. Q. Jin} \affiliation{%
Institute of Physics, Chinese Academy of Sciences - Beijing, China
}%
\author{M. Nicklas} \email{nicklas@cpfs.mpg.de}
 \affiliation{%
Max Planck Institute for Chemical Physics of Solids, 01187 Dresden, Germany
}%

\date{\today}

\begin{abstract}
Electrical-resistivity and magnetic-susceptibility measurements under hydrostatic pressure up to $p
\approx 2.75$~GPa have been performed on superconducting LiFeP. A broad superconducting (SC) region
exists in the temperature - pressure ($T-p$) phase diagram. No indications for a spin-density-wave
transition have been found, but an enhanced resistivity coefficient at low pressures hints at the
presence of magnetic fluctuations. Our results show that the superconducting state in LiFeP is more
robust than in the isostructural and isoelectronic LiFeAs. We suggest that this finding is related to
the nearly regular [FeP$_4$] tetrahedron in LiFeP.
\end{abstract}


\maketitle

The recently discovered iron-based superconductors attract a great deal of interest because of their
high critical temperatures up to $T_c=55$~K.
\cite{Kamihara06,Watanabe07,Kamihara08,Chen08,Chen081,Ren08,Sefat08} Soon after the discovery of
superconductivity in the iron and nickel based oxyphosphides, LaFePO \cite{Kamihara06} and
LaNiPO,\cite{Watanabe07} superconductivity was found in LaFeAsO$_{0.89}$F$_{0.11}$ (``1111" type)
with a critical temperature of about 26~K. \cite{Kamihara08} Furthermore, the application of
hydrostatic pressure leads to an increase of $T_c$ up to 43~K at about 4~GPa.\cite{Takahashi08} The
superconductivity in iron-pnictide compounds is closely related to their layered structure, where the
iron-pnictide layers are interlaced with charge reservoir layers. Electron or hole doping, both
inside and outside of the iron-pnictide layers, strongly affects the superconducting properties.

The effect of external pressure on the structural and electronic properties of the iron-based
superconductors can be subtle. In La(O$_{1-x}$F$_x$)FeAs and Sm(O$_{1-x}$F$_x$)FeAs the application
of pressure revealed an anisotropic lattice compressibility at low pressures, \cite{Takahashi09}
which results in a significant modification of electronic density of states (DOS). In optimally doped
La(O$_{1-x}$F$_x$)FeAs  $T_c$ decreases linearly with increasing pressure up to 30~GPa. This decrease
is accompanied by the lattice properties becoming less anisotropic.\cite{Garbarino08} The close
connection between structural properties and superconductivity is further shown in {\it
Re}FeAsO$_{1-x}$ ({\it Re} = rare-earth metal). Here, $T_c$ attains its maximum value where the
[FeAs$_4$] units form a regular tetrahedron.\cite{Lee08,Zhao08} In
Ba(Fe$_{0.92}$Co$_{0.08}$)$_2$As$_2$ (``122" type) the uniaxial pressure dependencies of $T_c$ are
highly anisotropic and quite pronounced.\cite{Hardy09} $T_c$ is anticipated to increase with
increasing $c/a$ ratio.

Superconductivity was reported in the ``111"-type materials LiFeAs \cite{Wang08a,Tapp08,Pitcher08}
and NaFeAs.\cite{Parker09} In contrast to the ``1111" and ``122" compounds and to the isostructural
NaFeAs no signature of a spin-density-wave (SDW) or structural transition has been observed in LiFeAs
regardless of having a similar charge density in the FeAs
layers.\cite{Clarina08,Goldman08,Jesche08,Rotter08} Recently, Deng {\it et al.}\ discovered
superconductivity below 6~K in the As free ``111" compound LiFeP,\cite{Deng09} which is isostructural
and isoelectronic to LiFeAs and can be considered as compressed LiFeAs. The occurrence of bulk
superconductivity in both stoichiometric LiFeAs {\it and} LiFeP makes them special among the
iron-pnictide materials. So far bulk superconductivity in a stoichiometric member of the
iron-arsenides and its isostructural phosphorous homolog has not been reported to the best of our
knowledge. External pressure and isoelectronic chemical substitution have a different effect on the
crystal structure.\cite{Deng09,Mito09} This allows for a detailed study of the influence of
structural properties on superconductivity. In this paper we study the effect of hydrostatic pressure
on LiFeP by electrical-resistivity $(\rho)$ and magnetic-susceptibility $(\chi_{\rm AC})$
experiments.


LiFeP polycrystalls were synthesized as described in Deng {\it et al.} \cite{Deng09}. We carried out
four-probe electrical-resistivity and AC-susceptibility measurements under hydrostatic pressure using
a physical property measurement system (PPMS, Quantum Design) and a commercial flow cryostat,
respectively, utilizing a LR700 resistance/mutual inductance bridge (Linear Research). A compensated
coil system placed outside of the pressure cell was used for the AC-susceptibility experiments.
Pressures up to 2.75~GPa were generated using a double-layer piston-cylinder type pressure cell.
Silicone fluid served as pressure transmitting medium. The pressure was determined at low
temperatures by monitoring the pressure-induced shift of the superconducting transition temperature
of lead placed close to the sample. The narrow width of the transition confirmed the good hydrostatic
pressure conditions inside the cell.

\begin{figure}[t!]
\includegraphics[angle=0,width=8.5cm,clip]{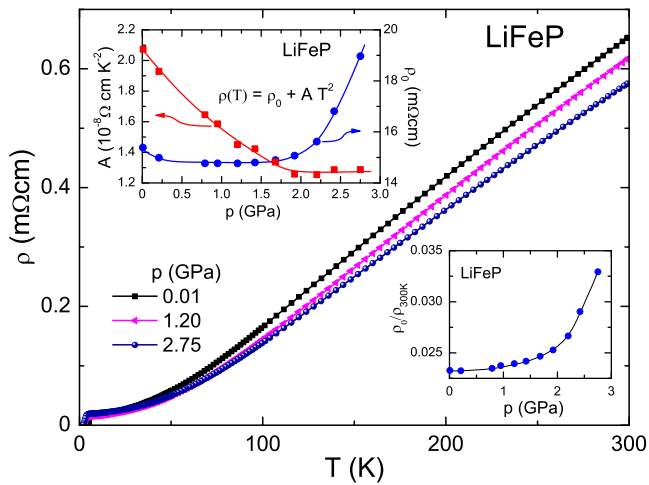}
\caption{\label{rho_highT} Electrical resistivity, $1.8{\rm~K}\leq T\leq 300{\rm~K}$, of LiFeP for
three representative pressures. The upper inset depicts the pressure dependence of the residual
resistivity $\rho_0$ and the prefactor $A$ obtained from a fit of $\rho(T)=\rho_0+AT^2$ to the
low-temperature normal-state resistivity. Details are given in the text. The lower inset displays the
pressure dependence of the ratio $\rho_0/\rho_{\rm 300\,K}$, where $\rho_{\rm 300\,K}$ is the
resistivity at $T=300$~K.}
\end{figure}

The temperature dependence of the electrical resistivity of LiFeP at three representative pressures
is depicted in Fig.~\ref{rho_highT}. In the normal state $\rho(T)$ exhibits a good metallic behavior
with no evidence for a SDW instability which is found in many of the ``1111"- or ``122"- type
iron-pnictide materials. A residual resistivity ratio $RRR=\rho_{\rm 300K}/\rho_0\approx43$ at
atmospheric pressure confirms the good quality of our polycrystalline sample. Here, $\rho_{\rm 300K}$
is the resistivity at 300~K and $\rho_0$ the residual resistivity. At low temperatures, a sharp
decrease of $\rho(T)$ to zero marks the onset of superconductivity, which is observed in the whole
investigated pressure range ($p\leq2.75$~GPa). The low-temperature normal-state resistivity follows a
$T^2$ dependence at all pressures indicating a Fermi-liquid state. The pressure dependence of the
parameters $\rho_0$ and $A$ of a $\rho(T)=\rho_0+A\,T^2$ fit to the data ($T_c\leq T\leq15$~K) is
presented in the upper inset of Fig.~\ref{rho_highT}. The observation of a $T^2$ behavior at such
elevated temperatures hints at the presence of strong electronic correlations. The temperature
coefficient $A$ is a measure of the quasiparticle - quasiparticle (QP - QP) scattering rate. $A(p)$
decreases by a factor of 1.6 from atmospheric pressure to $p=2$~GPa and stays constant with further
increasing pressure, indicating a reduction of the QP - QP scattering rate for $p\leq2$~GPa. The
enhanced QP - QP scattering rate at low pressures might be a hint for the presence of spin
fluctuations and indicate the proximity of LiFeP to magnetic order at ambient pressure despite no
direct evidence for long-range magnetic order has been found neither in LiFeP nor in its homolog
LiFeAs.

\begin{figure}[t!]
\includegraphics[angle=0,width=8.5cm,clip]{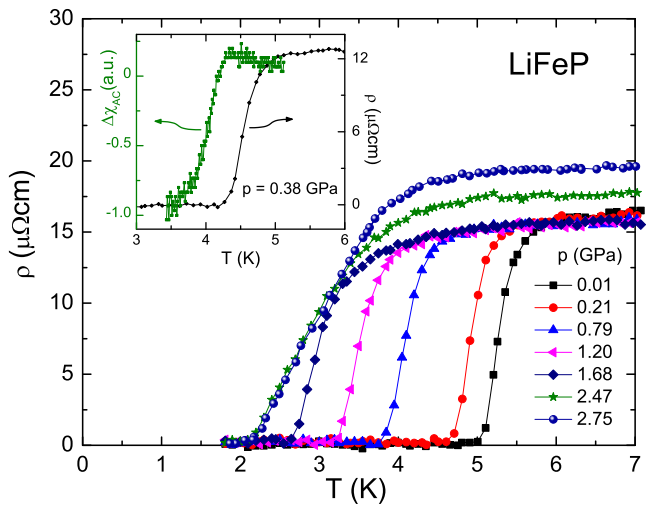}
\caption{\label{rho_lowT} Low-temperature electrical resistivity of LiFeP as function of temperature
for different pressures as indicated. The inset displays the AC susceptibility and the electrical
resistivity at $p=0.38$~GPa in the temperature region around the superconducting transition.}
\end{figure}

At ambient pressure, we find the onset of the resistive transition at about $\approx 6$~K in good
agreement with the literature.\cite{Deng09} Further on, we will use the $\rho(T)=0$ criterion to
define $T_c$ from our resistivity data. With increasing pressure the superconducting transition
shifts to lower temperatures (see Fig.~\ref{rho_lowT}). The width of the transition is nearly
pressure independent up to $p\approx 2.25$~GPa, even though the onset becomes more rounded before a
noticeable broadening becomes evident. The significant broadening is accompanied by an increase of
the low temperature normal-state resistivity, which is basically pressure independent below
$p\approx2$~GPa. This behavior is intrinsic to the sample and not caused by, e.g. cracks in the
sample, since the room-temperature resistivity, $\rho_{\rm 300K}(p)$, decreases monotonously upon
increasing pressure. This is also evidenced by the strong increase of the ratio $\rho_0(p)/\rho_{\rm
300K}(p)$ (see lower inset in Fig.~\ref{rho_highT}).

\begin{figure}[t!]
\includegraphics[angle=0,width=8.5cm,clip]{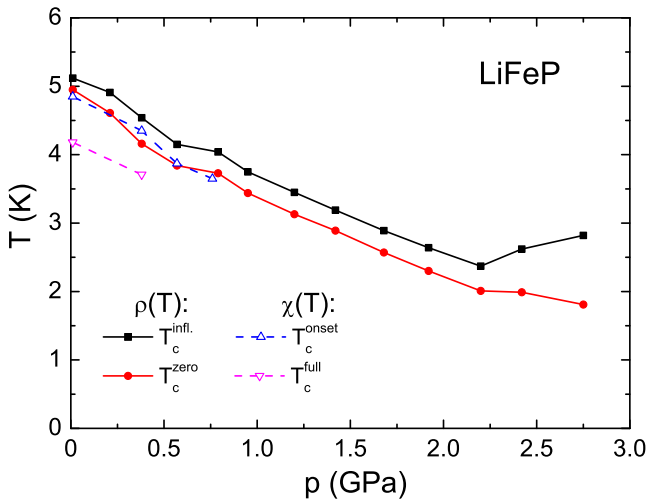}
\caption{\label{Tp_PhD} Temperature - pressure phase diagram of LiFeP. The solid symbols correspond
to results from $\rho(T)$ measurements. $T^{\rm infl.}_c$ is defined by the inflection point of
$\rho(T)$ and $T^{\rm zero}_c$ by the temperature where zero resistivity is obtained. The open
symbols correspond to $T_c$ determined by $\chi_{AC}(T)$ experiments. $T^{\rm onset}_c$ marks the
onset of the diamagnetic response and $T^{\rm full}_c$ the full transition.}
\end{figure}

In addition to $\rho(T)$ we measured $\chi_{\rm AC}(T)$ on the same sample and at the same pressures.
$\chi_{\rm AC}(T)$ exhibits a narrow, step-like feature at the superconducting transition. $\rho(T)$
reaches zero right at the temperature where $\chi_{\rm AC}(T)$ exhibits the onset of the diamagnetic
response. Above $p=0.79$~GPa, $T_c$ drops out of our measurement window for $\chi_{\rm AC}$.  The
inset of Fig.~\ref{rho_lowT} shows $\chi_{\rm AC}(T)$ and, for comparison, $\rho(T)$ at $p=0.38$~GPa.
The evolution of $T_c$ with increasing $p$ is depicted in Fig.~\ref{Tp_PhD}. The narrow width of the
superconducting transition in resistivity and, further, the good correspondence between $T_c$
determined by the $\rho(T)$ and the $\chi_{\rm AC}(T)$ in the $T-p$ phase diagram is unusual for
superconductivity in stoichiometric ``1111" and ``122" materials. There, quite often zero resistance
is found without any indication for bulk superconductivity or a very broad transition is observed
(e.g. Ref.~\onlinecite{Saha09,Kumar09}).

\begin{figure}[t!]
\includegraphics[angle=0,width=8.5cm,clip]{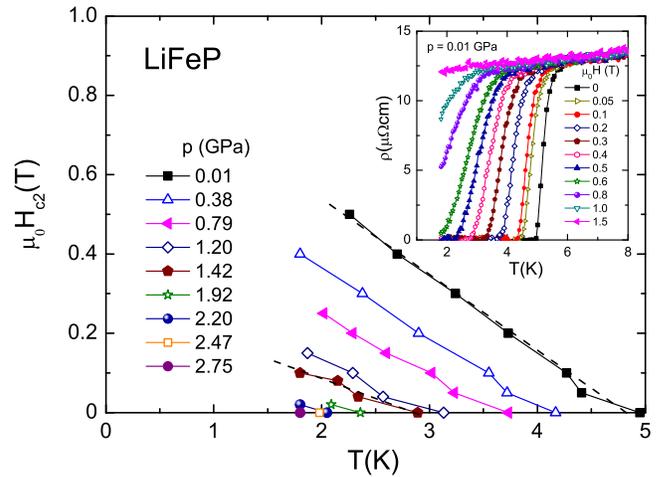}
\caption{\label{Hc2T} Magnetic field - temperature phase diagram of LiFeP for different pressures.
The dashed lines at $p=0.01$ and 1.42~GPa are serving as an example of the linear fits to the data.
The inset shows the resistivity data for $p=0.01$~GPa for different magnetic fields. The
zero-resistivity criterion was used for determining $T_c$.}
\end{figure}

To determine the superconducting upper-critical field, $H_{c2}(T)$, we conducted measurements of the
electrical resistivity in magnetic fields. $H_{c2}$ vs. $T$ curves at different pressures are
displayed in Fig.~\ref{Hc2T}. $H_{c2}(T)$ exhibits a roughly linear temperature dependence in the
accessible temperature range ($T\geq1.8$~K) with the exception of the first data point in magnetic
field ($\mu_0H=0.5$~T), which indicates the presence of a small tail. A similar tail has been
previously reported in other iron-based superconductors.\cite{Wang08,Hunte08,Miclea09} As possible
origin of the tail multiband effects were discussed. Increasing pressure suppresses $H_{c2}(T)$
effectively and, correspondingly, the absolute value of the slope $\mu_0{\rm d}H_{c2}(T)/{\rm d}T$ of
a straight-line fit to the data decreases from $1.92$~T/K at 0.01~GPa to $0.95$~T/K at 1.42~GPa.
Furthermore, with increasing magnetic field the superconducting transition in $\rho(T)$ gradually
broadens as shown for $p=0.01$~GPa in the inset of Fig.~\ref{Hc2T}. The broadening of the resistive
transition on increasing magnetic field indicates an anisotropy of $H_{c2}(T)$ as anticipated for a
quasi-two-dimensional electronic structure.\cite{Lebegue07}


In comparison with LiFeAs, LiFeP can be viewed as {\it compressed} LiFeAs. At $5.5-6.5$~GPa $T_c$ of
LiFeAs matches $T_c$ of LiFeP at atmospheric pressure: LiFeAs ``becomes"
LiFeP.\cite{Gooch09,Mito09,Zhang09} The lattice parameters obtained for LiFeP are $a =
3.692~{\rm\AA}$, $c = 6.031~{\rm\AA}$ \cite{Deng09} compared to $a = 3.670~{\rm\AA}$, $c =
6.108~{\rm\AA}$ for LiFeAs at $6.54$~GPa.\cite{Mito09} The lattice parameters $a$ and $c$ in LiFeAs
are contracted by 2.7\% and 3.9\%, respectively, at $6.54$~GPa, whereas the replacement of As by P
reveals a highly anisotropic contraction of $a$ and $c$ by 2.1\% and 5.1\%, respectively. This leads
to a smaller structural anisotropy in LiFeP compared to LiFeAs at $6.54$~GPa. It has been pointed out
for the iron-pnictides that $T_c$ attains maximum values when the [Fe{\it Pn}$_4$], where ${Pn}={\rm
P}$, As, form a regular tetrahedron.\cite{Lee08,Zhao08} At ambient pressure the [FeP$_4$] tetrahedron
of LiFeP is only slightly distorted with $\alpha = 108.58^\circ$ and $\beta =
109.92^\circ$,\cite{Deng09} while LiFeAs at $6.54$~GPa possesses a highly distorted tetrahedron
$\alpha = 99.39^\circ$ and $\beta = 114.70^\circ$.\cite{Mito09} The bond angle of a regular
tetrahedron is $109.47^{\circ}$. A nearly perfect [FeP$_4$] tetrahedron in LiFeP, but a highly
distorted [FeAs$_4$] tetrahedron in LiFeAs and taking into account a similar $T_c$ in both materials
suggest that the perfectness of the [Fe{\it Pn}$_4$] tetrahedron is not the determining property for
the value of $T_c$. Moreover, our result suggests that changes in the DOS other than those strictly
related to the perfectness of the [Fe{\it Pn}$_4$] tetrahedron are governing the value of $T_c$.
However, our experiments reveal that superconductivity in LiFeP is more robust than in LiFeAs. In
LiFeAs $T_c(p)$ decreases linearly on increasing pressure in the whole pressure range up to $\sim
10$~GPa.\cite{Zhang09} The initial slope of $T_c(p)$ $|{\rm d}T_c(p)/{\rm d}p\,|_{p=0}|=1.23$~K/GPa
for LiFeP is significantly smaller compared to the value in LiFeAs,  $|{\rm d}T_c(p)/{\rm
d}p\,|_{p=0}|=(1.56\sim2)$~K/GPa.\cite{Gooch09,Mito09} Since $T_c(p)$ decreases linearly in LiFeAs,
the same significant difference in the slopes of $T_c(p)$ is present when we compare them where the
$T_c$'s of LiFeP at $p=0$ and LiFeAs under pressure ($5.5-6.5$~GPa) are matching. This clearly
indicates that the superconductivity in LiFeP is more robust than in LiFeAs. This is furthermore
supported by a decreasing slope of $T_c(p)$ upon increasing pressure in LiFeP (see
Fig.~\ref{Tp_PhD}). Therefore, our study suggests that a more regular [Fe{\it Pn}$_4$] tetrahedron
``strengthens" the superconducting state, but is not determining the size of $T_c$.

We will now turn to the unusual increase of the low-temperature normal-state resistivity above
$p\approx2$~GPa. While the residual resistivity, $\rho_0(p)$, increases by about $1/3$ from 1.68~GPa
to 2.75~GPa, the $A$ coefficient stays nearly pressure independent in this pressure range. This
indicates that the QP-QP scattering rate does not change, but additional contributions to the
residual scattering appear and become stronger upon increasing pressure. Since, as we discussed
before, $\rho_{\rm 300K}(p)$ decreases in the mentioned pressure range and, thus, we can exclude an
extrinsic reason and, clearly, pressure does not add impurities, a different scattering mechanism has
to be considered. An increase of $\rho_0(p)$ is generally caused by additional disordered scattering
centers. A similar increase of the resistivity at low temperatures is observed in LiFeAs, but at much
higher pressures $p\gtrsim 11$~GPa.\cite{Zhang09} There, it has been proposed that additional
disordered scattering centers created by local magnetic ordering cause the enhanced
$\rho_0$.\cite{Zhang09} Increasing pressure reduces the in-plane Fe-Fe distance and concomitantly
enhances local magnetic correlations leading to additional magnetic scattering centers.


In summary, we have studied the $T-p$ phase diagram of the iron-pnictide superconductor LiFeP. Our
experiments evidence a more robust superconducting state than in the isostructural homolog LiFeAs. We
relate this to the nearly regular [Fe{\it Pn}$_4$] tetrahedron in LiFeP in contrast to the highly
distorted one in LiFeAs. However we do not find a general relationship of the bond angle $\alpha$ and
$T_c$ as suggested in literature.\cite{Lee08} Furthermore, we observe an enhanced QP-QP scattering
rate at low pressures, which might indicate the presence of spin fluctuations. However further
studies are needed to verify this speculation.



\begin{thebibliography}{ }


\bibitem{Kamihara06} Y. Kamihara, H. Hiramatsu, M. Hirano, R. Kawamura, H. Yanagi,
T. Kamiya, and H. Hosono, J. Am. Chem. Soc. {\bf 128}, 10012 (2006).

\bibitem{Watanabe07} T. Watanabe, H. Yanagi, T. Kamiya, Y. Kamihara, H. Hiramatsu,
M. Hirano, and H. Hosono,  Inorg. Chem. {\bf46}, 7719 (2007).

\bibitem{Kamihara08} Y. Kamihara, T. Watanabe, M. Hirano, and H. Hosono, J. Am.
Chem. Soc. {\bf 130}, 3296 (2008).

\bibitem{Chen08} X. H. Chen, T. Wu, G. Wu, R. H. Liu, H. Chen, and D. F. Fang,
Nature {\bf 453}, 761 (2008).

\bibitem{Chen081} G. F. Chen, Z. Li, D. Wu, G. Li, W. Z. Hu, J. Dong, P. Zheng,
J. L. Luo, and N. L. Wang, Phys. Rev. Lett. {\bf100}, 247002 (2008).

\bibitem{Ren08} Z. A. Ren, J. Yang, W. Lu, W. Yi, G. C. Che, X. L. Dong, L. L. Sun, and Z. X. Zhao, Mater. Sci.
Innov. {\bf12}, 105 (2008).

\bibitem{Sefat08} A. S. Sefat, M. A. McGuire, B. C. Sales,
R. Jin, J. Y. Howe, and D. Mandrus,  Phys. Rev. B {\bf77}, 174503 (2008).

\bibitem{Takahashi08} H. Takahashi, K. Igawa, K. Arii, Y. Kamihara,
M. Hirano, and H. Hosono, Nature {\bf 453}, 376 (2008).

\bibitem{Takahashi09} H. Takahashi, H. Okada, K. Igawa, Y. Kamihara, M. Hirano, H. Hosono, K. Matsubayashi, and Y.
Uwatoko, J. Supercond. Nov. Magn. {\bf22}, 595 (2009).

\bibitem{Garbarino08}  G. Garbarino, P. Toulemonde, M. \'{A}lvarez-Murga, A. Sow,
M. Mezouar, and M. N\'{u}\~{n}ez-Regueiro, Phys. Rev. B {\bf78},
100507(R) (2008).

\bibitem{Lee08} C.-H. Lee, A. Iyo, H. Eisaki, H. Kito, M. T. Fernandez-Diaz, T. Ito, K. Kihou, H. Matsuhata, M.
Braden, and K. Yamada, J. Phys. Soc. Jpn. {\bf77}, 083704 (2008).

\bibitem{Zhao08}  J. Zhao, L. Wang, D. Dong, Z. Liu, H. Liu, G. Chen, D. Wu, J. Luo, N. Wang, Y. Yu, C. Jin, and
Q. Guo, J. Am. Chem. Soc. {\bf130}, 13828 (2008).

\bibitem{Hardy09} F. Hardy, P. Adelmann, T. Wolf, H. v. L\"{o}hneysen, and C.
Meingast, Phys. Rev. Lett. {\bf102}, 187004 (2009).

\bibitem{Wang08a} X. C. Wang, Q. Q. Liu, Y. X. Lv, W. B. Gao, L. X. Yang, R. C. Yu,
F. Y. Li, and C. Q. Jin, Solid State Commun. {\bf148}, 538 (2008).

\bibitem{Pitcher08} M. J. Pitcher, D. R. Parker, P. Adamson, S. J. C. Herkelrath,
A. T. Boothroyd, R. M. Ibberson, M. Brunelli, and S. J. Clarke, Chem. Commun., 5918 (2008).

\bibitem{Tapp08} J. H. Tapp, Z. Tang, B. Lv, K. Sasmal, B. Lorenz, P. C. W. Chu,
and A. M. Guloy, Phys. Rev. B {\bf78}, 060505(R) (2008).

\bibitem{Parker09} D. R. Parker, M. J. Pitcher, P. J.
Baker, I. Franke, T. Lancaster, S. J. Blundell, and S. J. Clarke,
Chem. Commun., 2189 (2009).

\bibitem{Clarina08} C. de la Cruz, Q. Huang, J. W. Lynn, J. Y. Li, W.
Ratcliff II, J. L. Zarestky, H. A. Mook, G. F. Chen, J. L. Luo, N. L. Wang, and P. C. Dai, Nature
(London) {\bf453}, 899 (2008).

\bibitem{Goldman08} A. I. Goldman, D. N. Argyriou, B. Ouladdiaf, T. Chatterji, A.
Kreyssig, S. Nandi, N. Ni, S. L. Bud'ko, P. C. Canfield, and R.
J. McQueeney, Phys. Rev. B {\bf78}, 100506(R) (2008).

\bibitem{Jesche08} A. Jesche, N. Caroca-Canales, H. Rosner, H. Borrmann, A.
Ormeci, D. Kasinathan, H. H. Klauss, H. Luetkens, R. Khasanov, A.
Amato, A. Hoser, K. Kaneko, C. Krellner, and C. Geibel, Phys.
Rev. B {\bf78}, 180504(R) (2008).

\bibitem{Rotter08} M. Rotter, M. Tegel, D. Johrendt, I. Schellenberg, W. Hermes,
and R. P\"{o}ttgen, Phys. Rev. B {\bf78}, 020503(R) (2008).

\bibitem{Deng09} Z. Deng, X. C. Wang, Q. Q. Liu, S. J. Zhang, Y. X. Lv, J. L. Zhu,
R. C. Yu and C. Q. Jin, EPL {\bf87}, 37004 (2009).

\bibitem{Mito09} M. Mito, M. J. Pitcher, W. Crichton, G. Garbarino, P. J. Baker,
S. J. Blundell, P. Adamson, D. R. Parker, and S. J. Clarke, J. Am. Chem. Soc. {\bf131}, 2986 (2009).

\bibitem{Zhang09} S. J. Zhang, X. C. Wang, R. Sammynaiken, J. S. Tse, L. X. Yang,
Z. Li, Q. Q. Liu, S. Desgreniers, Y. Yao, H. Z. Liu, and C. Q. Jin, Phys. Rev. B {\bf80}, 014506
(2009).

\bibitem{Saha09} S. R. Saha, N. P. Butch, K. Kirshenbaum, and J. Paglione, and P. Y. Zavalij, Phys. Rev. Lett.
{\bf103}, 037005 (2009)

\bibitem{Kumar09}M. Kumar, M. Nicklas, A. Jesche, N. Caroca-Canales, M. Schmitt,
M. Hanfland, D. Kasinathan, U. Schwarz, H. Rosner, and C. Geibel, Phys. Rev. B {\bf78}, 184516
(2008).

\bibitem{Wang08} Z.-S. Wang, H.-Q. Luo, C. Ren, and H.-H. Wen, Phys. Rev. B {\bf78}, 140501(R)
(2008).

\bibitem{Hunte08} F. Hunte, J. Jaroszynski, A. Gurevich, D. C. Larbalestier, R. Jin, A. S. Sefat, M. A. McGuire, B. C. Sales,
D. K. Christen, and D. Mandrus, Nature {\bf 453}, 903 (2008).

\bibitem{Miclea09} C. F. Miclea, M. Nicklas, H. S. Jeevan, D. Kasinathan, Z. Hossain, H. Rosner, P. Gegenwart, C. Geibel, and
F. Steglich, Phys. Rev. B {\bf79}, 212509 (2009).

\bibitem{Lebegue07} S. Leb\`{e}gue, Phys. Rev. B {\bf75}, 035110
(2007).

\bibitem{Gooch09} M. Gooch, B. Lv, J. H. Tapp, Z. Tang, B.Lorenz, A. M. Guloy, and P. C. W. Chu, EPL {\bf85}, 27005 (2009).


\end{thebibliography}
\end{document}